\documentclass{ws-procs975x65}
\usepackage[utf8]{inputenc}
\usepackage{amsmath,amscd}
\usepackage{amsfonts}
\usepackage{amssymb}
\usepackage{graphicx,shortvrb}
\def\G{\Gamma}

\def\l{\lambda}

\def\m{\mu}
\def\n{\nu}

\def\r{\rho}

\def\s{\sigma}

\def\x{\xi}

\def\GN{\mathrm{G}_{\mathrm{N}}}

\def\tt{\tau}
\newcommand{\os}[2]{{\overset{\,\scalebox{0.5}{(#2)}}{#1}}{}}
\begin{document}

\title{Torsional Newton-Cartan gravity and strong gravitational fields}
\author{Dieter Van den Bleeken$^*$}

\address{Physics Department, Boğaziçi University,\\
Istanbul, Turkey\\
Institute for Theoretical Physics, KU Leuven\\
Leuven, Belgium\\
$^*$E-mail: dieter.van@boun.edu.tr}

\begin{abstract}
We review how the large $c$ expansion of General Relativity leads to an effective theory in the form of Twistless Torsional Newton-Cartan gravity. We show how this is a strong field expansion around the static sector of General Relativity and illustrate this through two examples. 
\end{abstract}


\bodymatter

%%%%%%%%%%%%%%%%% now a standard article style for the most part

\section{Introduction}
Weak field and/or nonrelativistic approximations of General Relativity (GR) have various applications, see e.g. Ref.~\refcite{PoissonW} for an introduction. These approximations are typically worked out in an adapted coordinate system. To the author's knowledge a covariant expansion of GR in the (inverse) speed of light $c$ was not systematically studied before the work of Dautcourt \cite{Dautcourt:1996pm} in the '90s. In that work it is assumed that the coefficient of $c^2$ in the relativistic connection vanishes and with this assumption the leading and subleading orders of the expansion reproduce Newtonian and post-Newtonian gravity respectively, see also Ref.~\refcite{Tichy:2011te}. Because the formalism is manifestly covariant one reproduces (post-)Newtonian gravity in a covariant formalism, so called Newton-Cartan (NC) gravity. The analysis of Ref.~\refcite{Dautcourt:1996pm} leaves however two questions: 1) why this assumption on the connection? 2) how come a large $c$ expansion automatically leads to a small $G_\mathrm{N}$ result? Indeed, as was recently shown\cite{VandenBleeken:2017rij}, these two questions are closely related: relaxing the assumption on the connection introduces (nonrelativistic) strong gravitational effects, in particular time dilation. Working out the expansion in full detail one obtains at leading order a generalization of Newtonian gravity, where this effect of time dilation is encoded as a torsion in the NC connection.\footnote{Since this talk was presented some further related work appeared\cite{Hansen:2018ofj,Cariglia:2018hyr}.}

\section{Review of the large $c$ expansion}
In the large $c$ expansion\cite{Dautcourt:1996pm,Tichy:2011te,VandenBleeken:2017rij} one starts from an expansion of the metric and its inverse in (even\footnote{It is consistent to consider only even powers of $c$, odd powers are allowed to appear however and including them is interesting\cite{vdbappear}.}) powers of $c^{-1}$:
\begin{equation}
g_{\m \n} = \sum_{i=-1}^\infty \os{g}{2$i$}\!_{\mu\nu} c^{-2i} \qquad g^{\m \n}= \sum_{i=0}^\infty\os{g}{2$i$}^{\m\n} c^{-2i}\label{metexp}
\end{equation}
The key physical input -- that time is related to space through a factor $c$ -- translates to the assumption that $\os{g}{-2}\!_{\mu\nu}$ is of rank 1 and negative, so we can write
\begin{equation}
\os{g}{-2}\!_{\mu\nu}=-\tau_\mu\tau_\nu\,\label{tt}
\end{equation}
One needs to impose the condition that $g_{\m\n}$ and $g^{\m\n}$ are each others inverse. For the leading term of the inverse metric this implies it is purely spatial, i.e. a rank three symmetric tensor orthogonal to the time direction:
\begin{equation}
\os{g}{0}^{\mu\nu}=h^{\m\n}\qquad\mbox{with}\qquad h^{\m\n}\tau_\nu=0\label{hdef}
\end{equation}
At subleading order one finds that\cite{VandenBleeken:2017rij}
\begin{equation}
	\os{g}{0}\!_{\m\n}=-2\hat \Phi \tau_\m\tau_\n+\hat{h}_{\m\n}\qquad
	\os{g}{2}^{\m\n}=- \hat\tau^\mu\hat\tau^\n+\hat\beta^{\m\n}\label{metex}
\end{equation}
where $ \tau_\nu\hat \tau^\nu+\hat h_{\m\r}h^{\r\n}=\delta^\n_\m\qquad \hat\tau^\r\hat\tau^\s\hat h_{\r\s}=0\qquad \tau_\m \hat{\beta}^{\m\n}=0$.

Expanding the metric compatibility equation for the relativistic covariant derivative naturally leads to the nonrelativistic connection\cite{VandenBleeken:2017rij}
\begin{equation}
\os{\G}{nc}_{\m\n}^\l=\frac{1}{2}h^{\l\r}\left(\partial_\m \hat h_{\r\n}+\partial_\n \hat h_{\m\r}-\partial_\r \hat h_{\m\n}+2\partial_\r \hat \Phi \tau_\m\tau_\n-4\hat{\Phi}(\tau_\m\partial_{[\n}\tau_{\r]}+\tau_\n\partial_{[\m}\tau_{\r]})\right)+\hat\tau^\l\partial_{\m}\tau_\n\label{connection}
\end{equation}
This connection is compatible with the Newton-Cartan structure, i.e. $\os{\nabla}{nc}_\mu h^{\nu\l}=0$ and $
\os{\nabla}{nc}_\mu\tau_\nu=0$, but has torsion $\os{\mathrm{T}}{nc}^\l_{\m\n}=2\hat\tau^\l\partial_{[\m}\tau_{\n]}$. 

One can now insert the expansion \eqref{metexp} into the relativistic Ricci tensor and expand it in turn in powers of $c^{-2}$. Equating each term to zero will provide an expansion of the vacuum Einstein equations\footnote{See Ref.~\refcite{VandenBleeken:2017rij} for the contributions of the energy momentum tensor.}. At leading order the Einstein equation is equivalent to
\begin{equation}
\tau_{[\mu}\partial_\n\tau_{\l]}=0\,.\label{TTeq}
\end{equation}
The geometric interpretation of the equation \eqref{TTeq} is two-fold: first it implies that $\tau_\mu$ defines a foliation and hence a global time direction, second it implies that the torsion of the connection \eqref{connection} is twistless\cite{Christensen:2013lma}, and the related geometry is referred to as Twistless Torsional Newton Cartan (TTNC) geometry. 

The next to leading order Einstein equation provides an equation for this torsion:
\begin{equation}
h^{\l\r}(\os{\nabla}{nc}_{\l}\hat a_\r-\hat a_\l\hat a_\r)=0\,,\qquad\mbox{where }\hat a_\mu={\cal L}_{\hat \tau} \tau_\mu\,. \label{aeq}
\end{equation}
At next to next to leading order the expanded relativistic Ricci tensor provides an equation for the Ricci tensor of the Newton-Cartan connection \eqref{connection}, see Ref.~\refcite{VandenBleeken:2017rij} for the actual expression, we'll here simply refer to it as the 'Ricci equation'. The component of this Ricci equation proportional to $\tau_\m\tau_\n$ provides the Poisson equation for the Newtonian potential $\Phi$. The Ricci equation together with \eqref{aeq} provide a system of equations for the fields determining the TTNC geometry and for this reason we'll refer to the theory formed by these equations as TTNC gravity.

In summary, up to next to next to leading order in a $c^{-2}$ expansion GR is effectively described by TTNC gravity.

\section{TTNC gravity as an expansion around static GR}
One of the defining features of NC geometry is that it is covariant under 4 dimensional diffeomorphisms. But since in TTNC there is a well defined time direction -- specified by $\tau_\mu$ -- on which all observers agree, this 4 dimensional diffeomorphism covariance is somewhat artificial. Indeed, from a physical point of view it might be more natural to introduce an explicit time coordinate $t$, such that $\tau_\mu=e^{\lambda(t,x)}\delta_\m^t$, the most general form allowed by \eqref{TTeq}. This choice of time breaks 4 dimensional diffeomorphisms to 3 dimensional time dependent diffeomorphisms.\footnote{See e.g. Ref.~\refcite{Bleeken:2015ykr} for some details on the precise relation between the 4d diffeomorphisms, Milne boosts and 3d time dependent diffeomorphisms.} It has the advantage that it removes a number of unphysical components of the fields and makes the physical field content more explicit. After this partial gauge fixing the expansion (\ref{tt}, \ref{hdef}, \ref{metex}) can be rewritten as
\begin{eqnarray}
ds^2&=&-c^2\, e^{\lambda} dt^2+e^{-\lambda}h_{ij}dx^idx^j-2e^\lambda dt(\Phi dt-C_idx^i)\nonumber\\
&&+c^{-2}\left((e^{-\lambda}\beta_{ij}-e^\lambda C_iC_j)dx^idx^j+B_idtdx^i+\Sigma dt^2\right)+{\cal O}(c^{-4})\label{startp}
\end{eqnarray}
The physical fields are then a scalar $\lambda$, a spatial 3-metric $h_{ij}$, the Newtonian potential $\Phi$,  a vectorfield $C_i$ and a symmetric 3-tensor $\beta_{ij}$ \footnote{Note that the fields $B_i$ and $\Sigma$ will not appear in the equations to the order of the expansion we consider in this work.}. This formulation makes manifest the connection between the NC torsion and the time-like warpfactor of GR, both determined by $\lambda$. Contrary to the Newtonian potential $\Phi$, which is the contribution to the warp factor at zeroth order the potential $\lambda$ appears at the higher order $c^2$, it encodes a strong gravitational effect that is absent in the weak field post-Minkowskian approximation.

The TTNC gravitational equations, \eqref{aeq} and the Ricci equation, take the following form in this formulation ($K_{ij}=2\partial_{[i}C_{j]}\,,\ G_i=-\partial_i\Phi-e^{-\frac{1}{2}\lambda}\partial_t(e^{\frac{1}{2}\lambda} C_i)$):
\begin{eqnarray}
\nabla_i\partial^i\lambda&=&0\qquad\qquad\qquad R_{ij}=\frac{1}{2}\partial_i\lambda\partial_j\lambda\label{stateq}\\
\frac{1}{2}\nabla^{j}\left(e^{2\lambda} K_{ij}\right)&=&-h^{jk}\nabla_{[j}\dot h_{i]k}-\frac{1}{2}(h^{jk}\dot h_{jk}-\dot{\lambda})\partial_i\lambda-\partial_i\dot{\lambda}+\dot h_{ij}\partial^j\lambda\label{Keq}\\
-e^{\frac{3}{2}\lambda}\nabla^i(e^{\frac{1}{2}\lambda}G_i)&=&T[\ddot\lambda,\ddot h,\partial\phi,\nabla\beta]\label{Peq}
\end{eqnarray}

The first important point to note is the hierarchical nature of the above equations. The first two equations  (\ref{stateq}) form an independent closed system of equations for $h_{ij}(t,x)$ and $\lambda(t,x)$ wich determines their spatial dependence but leaves the time dependence free. The third equation \eqref{Keq} is an equation that determines the spatial dependence of $C_i(t,x)$ and takes the time derivatives of $h_{ij}$ and $\lambda$ as input sources. The last equation \eqref{Peq} is an equation containing $\Delta\Phi$ and can thus be identified with the Newtonian Poisson equation, interestingly it contains second time derivatives of $h_{ij}$ and $\lambda$, and spatial derivatives of $C_i$ as sources\footnote{Note that we refrained from spelling out these source terms explicitely here. They can be obtained from Ref.~\refcite{VandenBleeken:2017rij} and as was pointed out there one can remove $\beta_{ij}$ from \eqref{Peq} by a gauge transformation. A better, gauge-independent, understanding has emerged from work\cite{Hansen:2018ofj} that appeared after this talk was presented: at one lower order there will appear an additional equation for $\beta_{ij}$ not including other new fields. Together with the equation \eqref{Peq} this will form a closed system determining the spatial dependence of $\Phi$ and $\beta_{ij}$.}.

Secondly one observes\footnote{This was pointed out by J. Raeymaekers.} that the equations \eqref{stateq} are essentially the static Einstein equations. Indeed any static metric can be written in the form $ds^2=-c^2\, e^{\tilde\lambda(x)} dt^2+e^{-\tilde\lambda(x)}\tilde h_{ij}(x)dx^idx^j$ and for such an ansatz the vacuum Einstein equations reduce to equations of the form \eqref{stateq}. The key difference between \eqref{stateq} and the static Einstein equations is that contrary to $\tilde \lambda$ and $\tilde h_{ij}$ the fields $\lambda$ and $h_{ij}$ are time-dependent. We come to the conclusion that $\lambda$ and $h_{ij}$ correspond to solutions of the static Einstein equations with any integration constants replaced by arbitrary functions of time. Such time-dependent deviations from static solutions then source the equations for the higher order fields, starting a series of corrections. One thus sees that the large $c$-expansion is an expansion around the static sector of GR. In particular any static solution\footnote{The example of the Schwarzschild black hole was worked out in Ref.~\refcite{VandenBleeken:2017rij} and extends to other examples such as Tolmann-Oppenheimer-Volkov fluid stars\cite{inprog}.} of GR will also be an exact (i.e. contrary to infinite series) solution of the effective TTNC gravity theory (by putting all fields other than $\lambda$, $h_{ij}$ to zero.) This makes very concrete how the large $c$-expansion is an extension of Newtonian gravity that also contains strong gravitational effects: indeed all the non-linearities of the static sector of GR are included.

\section{Non-static examples}
This interpretation of the large $c$-expansion as an expansion around the static sector of GR can best be explored in some non-static examples.
\subsection{Kerr black hole}
The Kerr black hole is a good starting point as it is no longer static but still stationary. The metric\footnote{We follow the notation of Ref.~\refcite{Wald:1984rg}.} contains two physical parameters $
M=m\GN c^{-2}\ a=J(mc)^{-1}=Ac^{-1}$. We consider the 'strong field'\cite{VandenBleeken:2017rij} expansion $c\rightarrow \infty$ with $M$ and $A$ fixed. Comparing with \eqref{startp} one finds
\begin{eqnarray}
\lambda&=&\log\left(1-\frac{2M}{r}\right)\qquad h_{ij}dx^idx^j=dr^2+\left(1-\frac{2M}{r}\right)r^2d\Omega^2\label{schw}\\
C_i dx^i&=&2 \frac{A M}{r} \sin ^2\theta\left(1-\frac{2M}{r}\right)^{-1}d\phi\qquad 
\Phi=\frac{A^2 M}{r^3}\left(1-\frac{2M}{r}\right)^{-1} \cos^2\theta
\end{eqnarray}
This example -- of which one can check that it solves (\ref{stateq}-\ref{Peq}) -- shows the Kerr black hole as a nonrelativistic expansion around the Schwarzschild black hole, which is a solution to \eqref{schw} \cite{VandenBleeken:2017rij}. The non-zero rotation of Kerr -- i.e. $A$ -- translates to a non-zero $C_\phi$ and $\Phi$. Note that the fields above only express the first terms in an infinite expansion of the Kerr metric. For these first terms to be a good approximation of the relativistic metric one needs the condition
\begin{equation}
r\gg r_++\frac{A^2}{Mc^2}+{\cal O}(c^{-4})\,,
\end{equation}
which means that at a distance $\frac{A^2}{Mc^2}$ from the outer horizon the nonrelativistic approximation breaks down. Interestingly the strong field expansion remains valid for large radii where gravity becomes weak and where it overlaps with the post-Newtonian expansion obtained by keeping $m$ and $J$ fixed as $c\rightarrow\infty$. This suggests that the TTNC effective theory is a resummation of the post-Newtonian series.

\subsection{Oppenheimer-Snyder collapse}
\begin{figure}
	\begin{center}
	\includegraphics[width=2in]{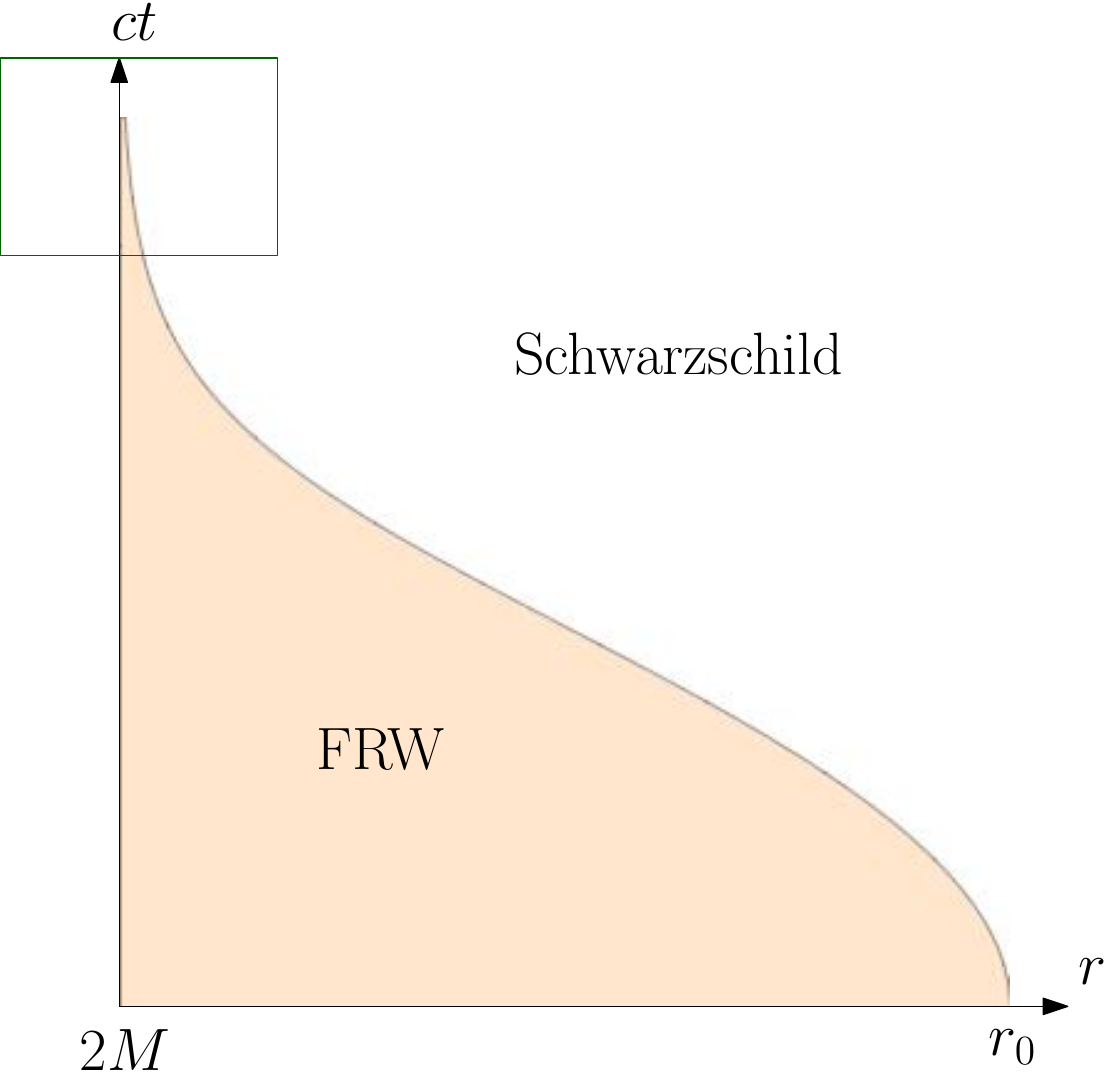}
	\caption{Space-time of a collapsing ball of dust.}
	\label{pic}
	\end{center}
\end{figure}

Realistic dynamic situations can be far from stationary, and so it is interesting to explore what additional effects can appear in their large $c$-expansion. To this aim we shortly present some aspects of the nonrelativistic expansion of the Oppenheimer-Snyder solution\cite{Oppenheimer:1939ue} which describes the collapse of a ball of dust. This solution depends on two parameters: $r_0$, the initial radius of the ball and $M=m\GN c^{-2}$, its total mass. It is composed of two metrics -- an FRW universe inside the ball and the Schwarzshild metric outside -- glued together along the worldtube of the ball's surface. This surface follows a radial geodesic in the Schwarschild geometry, see figure \ref{pic}, and can be described through the curve $r_\mathrm{match}(t)$, with $r$ and $t$ the standard Schwarzschild coordinates.

Again we'll be interested in a 'strong field'\cite{VandenBleeken:2017rij} expansion, where we keep $r_0$ and $M$ fixed as $c\rightarrow \infty$, and one finds
\begin{equation}
r_\mathrm{match}(t)\approx 2M\left(1+2\sqrt{\frac{r_0}{2M}-1}\,e^{-\frac{ct}{2M}}+{\cal O}(e^{-\frac{ct}{M}})\right)\label{matchapprox}
\end{equation}
Note that the expansion here is of a very different type than the standard Newtonian expansion which holds far away from the ball of dust, just after the collapse has started. In \eqref{matchapprox} we are close to the event where the ball forms a horizon, both in space and time, the green box in figure \ref{pic}. There gravity is strong and non-Newtonian, still it is nonrelativistic. That this is the case follows from the well-known -- but ever intriguing -- fact that an outside observer never sees a horizon form. For such an observer the radial velocity of the shell goes asymptotically to zero again at late times, making it describable by non-relativistic physics. Interestingly though we see that the form of the first correction in \eqref{matchapprox} is non-perturbative in large $c$. It appears that to describe the nonrelativistic gravitational fields of such a collapsing ball one would need a transseries extension of the ansatz \eqref{metexp}.

\section*{Acknowledgments}
The author thanks J. Hartong, J. Raeymaekers, and B. Vercnocke for useful discussions and comments. This work was partially supported by the Boğaziçi  University  Research  Fund  under  grant  number  17B03P1 and TUBITAK grant 117F376.

\end{document}